\documentclass[ ]{copernicus2}

\usepackage{color}
\usepackage[T1]{fontenc}

\begin{document}

\title{Electron cyclotron maser instability (ECMI) in strong magnetic guide field reconnection 
}

\author[1]{R. A. Treumann\thanks{Visiting the International Space Science Institute ISSI, Bern, Switzerland  \\ \\ \emph{Correspondence to}: R. A. Treumann (treumannr@gmail.com)}}
\author[2]{W. Baumjohann}

\affil[1]{Department of Geophysics and Environmental Sciences, Munich University, Munich, Germany}
\affil[2]{Space Research Institute, Austrian Academy of Sciences, Graz, Austria}

\runningtitle{Reconnection as ECMI radiation source}

\runningauthor{R. A. Treumann and W. Baumjohann}

\received{ }
\revised{ }
\accepted{ }
\published{ }


\firstpage{1}

\maketitle

\section*{Abstract}
The ECMI model of electromagnetic radiation from electron holes is shown to be applicable to spontaneous magnetic reconnection. We apply it to reconnection in strong current-aligned magnetic guide fields. Such guide fields participate only passively in reconnection, which occurs in the antiparallel components to both sides of the guide-field-aligned current sheets with current carried by kinetic Alfv\'en waves. Reconnection generates long (the order of hundreds of electron inertial scales) electron exhaust regions at the reconnection site {\sf X}-point, which are extended perpendicular to the current and the guide fields. Exhausts contain a strongly density depleted hot electron component and have properties similar to electron holes. Exhaust electron momentum space distributions are highly deformed, exhibiting steep gradients transverse to both the reconnecting and guide fields. Such properties suggest application of the ECMI mechanism with the fundamental ECMI X-mode emission beneath the nonrelativistic guide-field cyclotron frequency in localized soure regions. An outline of the mechanism and its prospects is given. Potential applications are AKR in auroral physics, solar radio emissions during flares, planetary emissions and astrophysical scenarios (radiation from stars and compact objects) involving the presence of strong magnetic fields and field-aligned currents. Drift of the exhausts along the guide field maps the local field and plasma properties. Escape of radiation from the exhaust and radiation source region still poses a problem. The mechanism can be studied in 2-D particle simulations of strong guide field reconnection which favours 2-D, mapping the deformation of the electron distribution perpendicular to the guide field, and using it in the numerical calculation of the ECMI growth rate. The mechanism suggests also that reconnection in general may become a source of the ECMI with or without guide fields. This is of particular interest in extended turbulent plasmas where reconnection serves as an integral dissipation mechanism of turbulent energy in myriads of small-scale current filaments. 

\vspace{0.5cm}
\section{Introduction}
The present note is intended to present a scenario, not a fully developed theory, for the generation of intense radio radiation in a strongly magnetized plasma by the Electron Cyclotron Maser Instability (ECMI) mechanism. This mechanism is a non-thermal source of radiation in radio waves which we do not review here in detail \citep[cf., e.g.,][for a still not outdated review]{treumann2006}. In the recent past it has experienced a few particular but no fundamental specifications \citep[as for recent examples one may consult][]{zhao2016,tong2017} respectively complications (inclusion of electron beams, calculation of low harmonics, numerically obtained graphs of growth rates etc.). The new scenario contrasts these attempts inasfar as it opens up a new direction in ECMI research. We don't do any calculation because the problem is too complex. A quantitative treatment requires a full-particle simulation including radiation. This will become clear when presenting the idea and describe the mechanism. The idea and scenario are interesting enough for injection and stimulation of a quantitative treatment by researchers working in the two sectors, collisionless reconnection (by numerical PIC simulations) and generation of electromagnetic radiation (using the properties inferred from the PIC simulations in a numerical calculation of the ECMI growth rates).

\subsection{Review and justification}
In order to provide a brief justification for attempting a new twist in ECMI theory we just mention that generation of high-intensity free space waves in a plasma, though for prospective reasons highly desirable from the geophysical, space and astrophysical viewpoint, poses a major problem. Any escaping radiation is based on the dynamics of electrons. In the electron equation of motion it is bound to the third time derivative$\vec{\cdots}\atop{}$ {\hspace{-2.45ex}$\vec{v}$} of the electron velocity $\vec{v}$. Gyro and synchrotron radiation is thus weak (providing very weak energy losses only), though it is comparably easy to treat \citep[see, e.g.,][]{jackson1962} while, when detected, maps temperature, magnetic field strength and mean plasma density \citep[see, e.g.,][]{rybicki1979}. Guided by the whistler instability, nonthermal radiation in the free space modes X and O from hot plasmas has early on provoked attempts to base it on thermal anisotropies in the electron distribution \citep{twiss1958a,hirshfield1963,melrose1976,melrose1978}. These anisotropies, however, turned out unreasonably large for causing instability, provoking doubt in this mechanism \citep[cf., e.g.,][]{melrose1978} compared to synchrotron radiation. The breakthrough \citep[see][]{wu1979}, following the earlier suggestions   \citep{twiss1958,hirshfield1964,melrose1976}, was brought by the observation that it is the relativistic effect in the electron distribution, if properly accounted for in the resonance curve, which must be added in order to obtain positive growth rates and generate high radiation intensities. Positive perpendicular-momentum space gradients in the weakly-relativistic electron distribution turned out to be the main \emph{necessary} condition for instability. Radiation in the fundamental X-mode was found beneath but close to the local (non-relativistic) electron gyrofrequency $\omega_{c}= eB/m$. Its excitation requires, in addition, as a \emph{sufficient} condition, that the plasma density is low, specifically the ratio of plasma-to-gyrofrequency $(\omega_e/\omega_c)^2\ll1$ must be sufficiently low for reaching substantial growth rates and, hence, radiation intensities. 

 The simple physics behind the ECMI is that, at such low densities, plasmas are unable to absorb the unstably generated waves. There is not enough plasma to digest the extra free energy. As a matter of fact, the local absorption coeffcient turns negative, and the whole plasma region becomes an emitter of radiation, mainly in the X-mode and weaker in the O-mode. There is, of course, also excitation of the Z-mode below the upper hybrid frequency \citep[cf., e.g.,][]{yi2013}, but this is basically electrostatic and cannot leave the plasma without invoking other wave transformation processes. 

\subsection{Modes, harmonics, thermal levels}
\begin{figure}[t!]
\includegraphics[width=0.5\textwidth,clip=]{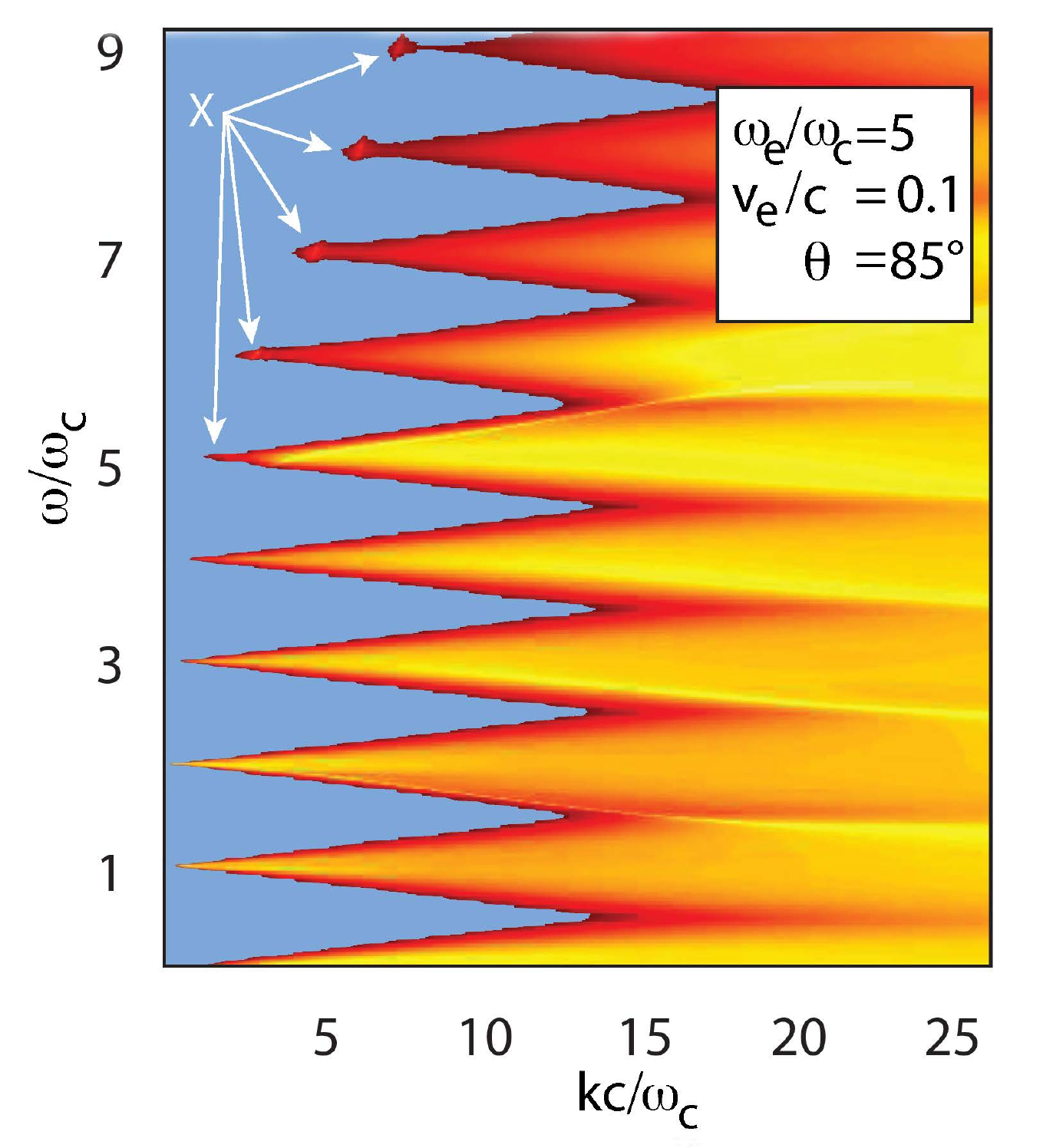} 
\caption{The thermal spectral densities of electromagnetic waves in a magnetized plasma for nearly perpendicular propagation \citep[after][]{yoon2017} for an electron plasma-to-cyclotron frequency ratio of $\omega_e/\omega_c=5$. The thermal level is structured by the electromagnetic electron cyclotron harmonics (Bernstein modes). The red intensified spots in the thermal noise starting at $\ell=5$ are located along a parabola of rising frequency, which is  the free-space branch of the X-mode in combination with the electromagnetic Bernstein modes. They may serve as cyclotron harmonic seed electromagnetic fluctuations for the X mode which, when becoming amplified by the ECMI, can propagate and escape from the plasma.}\label{fig-ecmi-one}
\end{figure}

The most general expression of the ECMI growth rate for the X-mode can be found in \citet{melrose1986} for arbitrary weakly relativistic electron distributions {\citep[cf., also][for application to AKR]{pritchett1986,pritchett1999}}. In principle it gives the growth rate $\Im(\omega_\ell)$ of the ECMI for any electron cyclotron harmonic $\ell=1,2,3\dots$ of $\omega_\ell\sim \ell\omega_c$. Inspection shows that, to leading order in $\ell$, the growth rate \emph{decreases} with order as $\Im(\omega_\ell)\sim \ell^{-1}$. Observations of the X-mode, mostly in the auroral region, do indeed indicate the presence of low harmonics \citep[as was noted by][and others]{pritchett1999,treumann2011}.\footnote{{Any wave must grow from its thermal level. In view of the fact that the thermal level of the fundamental X mode is very low (see Fig. \ref{fig-ecmi-one}) and its restricted possibilities of escape, it becomes questionable whether the observed AKR is indeed propagating on the fundamental. It may be probable that it propagates on one of the lower cyclotron harmonics, because many observations locate its frequency at 300-700 kHz which for the fundamental üposes the source rather deep into the upper auroral ionosphere.}} In fact, for those harmonics, if excited sufficiently strongly, the above necessary condition is obsolete as higher harmonics can propagate out and can be observed from remote. One may note that AKR radiation observed from remote is possibly not in the fundamental but harmonic putting its source a bit higher in the upper auroral ionosphere.

Excitation of any wave by instability requires the presence of a (weak) thermal background fluctuation at the unstable frequency from which it can grow \citep[for the kinetic theory of electromagnetic fluctuations see, e.g.,][]{schlickeiser2014a,schlickeiser2014b,schlickeiser2015}. Figure \ref{fig-ecmi-one} \citep[after][]{yoon2017} shows that the electromagnetic thermal fluctuation spectrum (for almost perpendicular propagation) in a thermal magnetized plasma is structured harmonically by electromagnetic Bernstein modes. These fluctuations intensify at long wavelengths (small wavenumbers $\vec{k}$) along the propagating X-mode dispersion (for the parameters chosen in this case starting at the fifth harmonic). Hence, since the ECMI unstably excites higher harmonics in the X mode, these encounter an appropriate thermal level from which they can grow. Note, however, that the thermal level for the fundamental ($\ell = 5$ in this case) is rather weak compared to the next low harmonics, implying that low harmonics will initially grow faster than the fundamental. On the other hand, if the ECMI would be able to also excite electromagnetic Bernstein modes, then these modes would find plenty of thermal fluctuations available which they could amplify to become free space harmonically structured propagating modes at shorter than X mode wavelengths. A calculation like this one is still outstanding but might be worth to perform in view of the broad prospects of applications (for instance in explanation of harmonic emissions during solar radio events, like in type I and type IV radio bursts, fiber bursts etc., possibly even radio emission from Jupiter etc.).  

The most interesting mode for the purposes of the following model, {which does not exclude the possible excitation of any of the higher cyclotron harmonics,} is the (fundamental) X-mode. The fundamental cannot leave the plasma, because if driven unstable by the ECMI it propagates at frequency $\omega\lesssim\omega_c$ \emph{below} its high-frequency cutoff on the lower (trapped) X-mode branch. Thus, escape of the fundamental mode to become visible requires propagation out of the source region to lower magnetic field strengths.This is particulary true for the most famous region of excitation of the ECMI, the auroral upper ionosphere/lower magnetosphere of Earth ({where it is the typical case}). In order to have the wanted efficiency of the mechanism, complicate if not exotic forms of the electron distribution need to be imposed. Among those a variety has been investigated in the literature, with global loss-cone distributions \citep[cf., e.g.][for elaborate examples]{louarn2006} of steep perpendicular-momentum-space gradients traditionally having been favoured. 

\subsection{Localized sources: electron hole model}
In some more recent papers \citep{treumann2011,treumann2012} we proposed that \emph{electron holes}, generated by strong field-aligned currents in a magnetized plasma, could possibly become sources of the ECMI. The existence of electron holes in both the downward current region \citep[cf., e.g.][]{carlson1998,ergun1998,ergun2002} and in the upward current region \citep{pottelette2004,pottelette2005} has been observationally confirmed, with the abundance of such holes in the downward current region being overwhelmingly higher. This electron hole model of the ECMI has the (technical) advantage that the relativistic resonance curve in momentum space becomes a circular section, a fact that substantially simplifies the calculation. In addition, it reproduces the striking narrowness of the emissions in frequency space. Moreover, the Doppler shift of the emissions caused by the motion and dynamics of the electron holes nicely reproduces the observed modulations in the narrow electron cyclotron emission lines. 

Excitation in the downward current region, where the strongest auroral-zone field-aligned currents flow and electron holes are ubiquitous, directly explains the observation of strong radiation in the upward current region by the simplicity by which the radiation can escape from the downward to the upward current region across the steep transverse plasma gradient between adjacent two of them. 

However, the great and so far unsurmountable \emph{disadvantage} of that model is, that electron holes are Debye-length objects on scales $\lambda_D=v_e/\omega_e$ (with $v_e$ the electron thermal speed). Such spatial structures should barely be able to excite strong radiation at several km-wavelengths as they are simply too small. One would need to deal with a spatial distribution of very many such holes located in a volume over one radiation wavelength in order to determine their collective (overlap of a series of holes by a much longer unstable wavelengths) and statistical (contribution of very many holes distributed in the volume to wave excitation) effects which might smear out most of the fine structure. Such a calculation, though absolutely worth of being applied in the electron hole radiation mechanism, has not yet been performed until today.

\subsection{Requirements on an efficient localized source}
In order to cure the deficiency of the short scale, is becomes necessary to look for other desirably common and possibly more frequent reasons of 
\begin{itemize}
\item[$(a)$] generation of electron depletions of sufficiently large spatial extension $L\gg\lambda_D$, \vspace{-1ex}
\item[$(b)$] these regions should contain sufficiently anisotropic or otherwise strongly deformed,  just weakly relativistic electron momentum distributions in\vspace{-1ex}
\item[$(c)$] sufficiently strong magnetic fields, thereby maintaining the advantages of the electron hole model.
\end{itemize}

\section{Collisionless reconnection: relevance for the ECMI}
A most promising candidate mechanism of this kind is collisionless reconnection which is well known to locally generate regions of vastly (sometimes in the simulations up to $\gtrsim90\%$) depleted electron densities, containing strong electric potential drops which cause electron heating and acceleration, and generate violently deformed electron-momentum-space distribution functions. 

\begin{figure}[t!]
\includegraphics[width=0.5\textwidth,clip=]{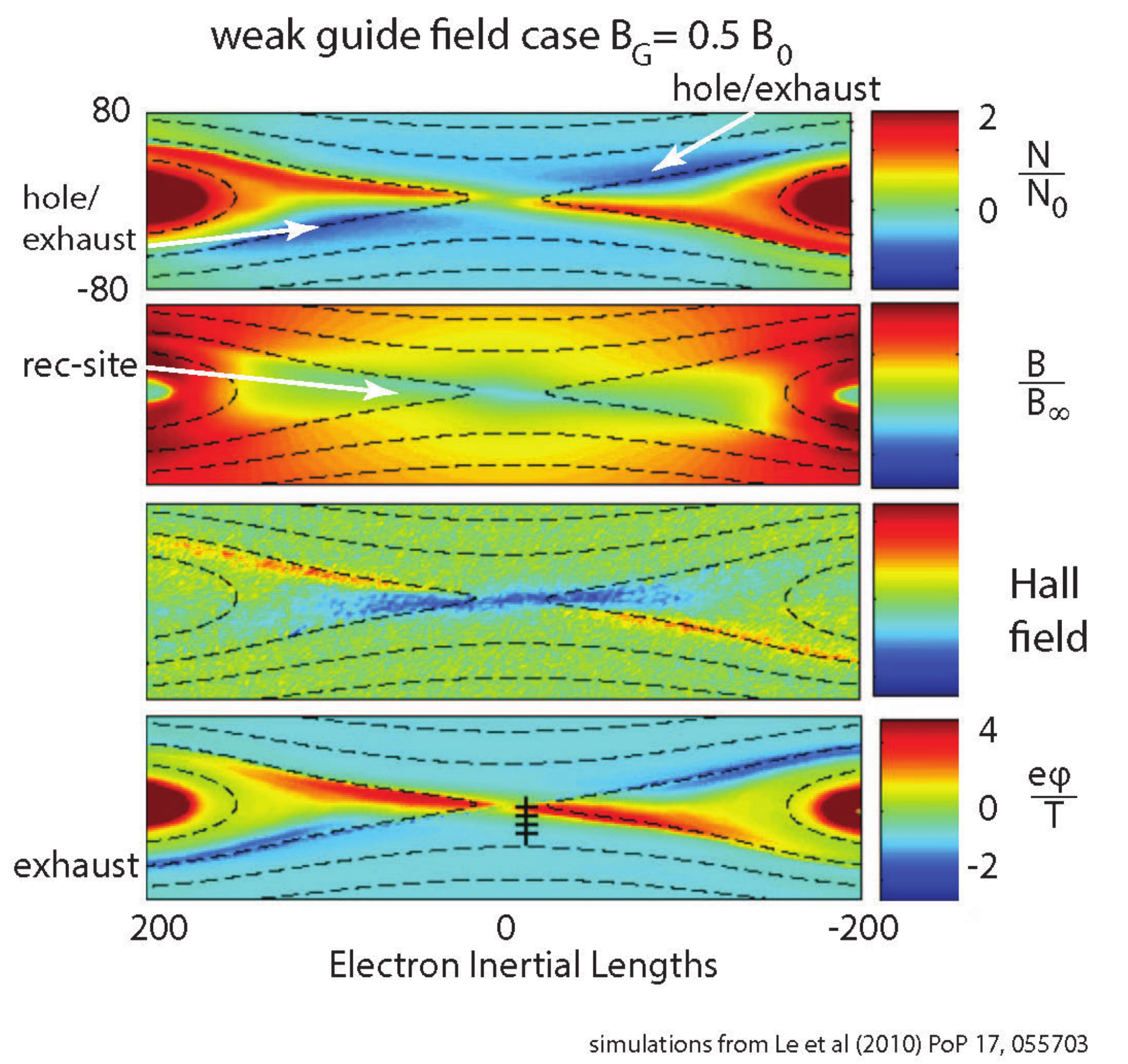} 
\caption{Weak-guide ($B_g=0.5 B_0$) field PIC-simulations of collisionless spontaneous reconnection in 2d \citep[after][]{le2010} with current and guide field perpendicular to the plane. From top to bottom normalized density, normalized magnetic field, Hall field structure, electric reconnection potential. The density exhibits the asymmetric exhaust (deep blue) and (red) electron accumulation regions caused by the $E\times B_g$ drift in the reconnection electric and guide fields. The nearly rectangular box reconnection region centred at the $\sf{X}$ point is characterized by a weak magnetic field. Very weak Hall fields arise due to the immobile ion background which for our purposes are of no interest. They just contribute to asymmetry. The electric potential shows region of different signs equivalent to the density structure. Note that in this  writing positive potentials attract, negative potentials expel electrons. Note the very long horizontal extension of the reconnection site which reaches a length of roughly $L\sim 300\;\lambda_e$ which means $\pm150\;\lambda_e$ to both sides in spite of the fact that the electrons in the guide field remain partially magnetized. The ion inertial length $\lambda_i\approx 43\;\lambda_e$ implies that the reconnection site is elongated to on each side of  up to $\gtrsim3 \lambda_i$. The magnetization of the electrons in the guide field perpendicular to the reconnection field affects the reconnection process very little only.} \label{fig-ecmi-two}
\end{figure}

Reconnection has so far not been considered yet in relation to the ECMI. However, a large number of kinetic-numerical PIC (particle-in-cell) simulations of collisionless reconnection have in the past three decades been put forward in two (2-D) and recently also in three (3-D) dimensions. Therefore, the relevant physics (still excluding many details) of collisionless reconnection under various different settings is, by now, quite well understood, mainly under conditions when two magnetic fields of equal (or not vastly different) strengths and mutually moderately inclined directions collide, merge and annihilate their antiparallel components, i.e. undergo reconnection \citep[for recent reviews of related aspects of collisionless reconnection cf., e.g.,][]{treumann2013,treumann2015}. In this process the energy of the annihilated fields goes into the electric potential which causes the mentioned electron depletions, heating, and acceleration. [For more general aspects of reconnection, including collisions and laboratory effects, the reader is referred to \citet{zweibel2009} and \citet{yamada2010}.] 

In the following, when talking about reconnection, we mean \emph{collisionless reconnection} as this is the only form of reconnection which is relevant on the scales under consideration. From the point of view of the physics of hot diluted plasmas it is highly questionable whether any of the originally proposed collisional (magnetohydrodynamic) versions of reconnection is realized anywhere else in nature other than in the deep molten metallic interiors of planets, where resistive reconnection participates in the magnetic dynamo action, and the highly compressed resistive stellar convection zones above the cores, where it participates as an important energy dissipation process in nuclear fusion. 

\subsection{Reconnection as generator of electron quasi-holes}
So far reconnection has been investigated mainly in counterstreaming flows/colliding magnetic fields under small angles and for plasma-$\beta=2\mu_0NT_e/B_0^2>1$. Under such conditions the magnetic field is weak and, though diluted, the surrounding ambient plasma remains superdense: $\omega_e>\omega_c$. Clearly the ECMI, even if excited in the fundamental X mode at the reconnection site, will ($a$) be of very weak intensity, because the dissipated magnetic energy is comparably small, and ($b$) will -- similarly to the electron hole model -- be trapped in the local density depletions near the $\mathsf{X}$-point, thus unable to escape. This case is of no interest here though, as noted above, at higher harmonics the underdensity condition is of lesser importance. One may thus expect that reconnection sites, even under weak field conditions, may show some radiative signatures in the low-electron-cyclotron harmonics thereby mapping the magnetic field strength around the reconnection sites.\footnote{Recognizing this possibility is of substantial interest in astrophysics as a mechanism that may cause a large volume of highly turbulent plasma to become glowing in radio emission from the myriads of small-scale reconnecting current filaments undergoing spontaneous reconnection \citep{treumann2015}. Such a radio glow is conventionally interpreted as sychrotron radiation but may indicate something completely different: turbulence, magnetic field strength, plasma density obtained from the electron inertial scale, reconnection and dissipation of turbulent energy in the many small-scale current filaments. Though the dissipation mechanism has not yet been investigated in detail, its clarification would in this case provide important information about the matter involved.}

An example of a 2-D \emph{weak guide-field} simulation of reconnection is depicted in Fig. \ref{fig-ecmi-two} for the case of a guide field $B_g=\frac{1}{2}B_0$ half the value of the reconnection field $B_0$. The electron density (first panel) is seen to be asymmetric with respect to the central $\mathsf{X}$-point where the magnetic fields $\pm B_0$ from above and below come into contact and reconnect, with magnetic field strength strongly reduced (second panel). The density near the reconnection site in the current centre exhibits two regions of density increase (red) and two regions (dark blue) where the electrons are depleted, indicated as \emph{exhausts}. These exhausts do strongly resemble the main property of electron holes which are a depletion of the electron density on the Debye scale. \emph{Electron exhausts thus represent substantially more extended electron quasi-holes}.

\subsection{Exhaust properties of relevance for the ECMI 1}
The exhaust length is of the order of $\gtrsim 100\: \lambda_e$ or more, with $\lambda_e=c/\omega_e$ the electron inertial length, and width the order of $20\: \lambda_e$. The ratio of electron inertial to Debye length is $\lambda_e/\lambda_D=c/v_e\gtrsim 10\gg 1$, making the size of the exhausts substantially larger, roughly a factor of several 100, than the size of an electron hole in both directions perpendicular to the ambient (guide) field. 

A large size like this one is clearly advantageous in view of the ECMI and its wavelength. In fact, since $\lambda_e\sim 1$ km or so in the auroral ionosphere, the size of the exhaust allows for several tens of km-length waves to propagate inside the exhaust (as weak thermal fluctuations ready for amplification). This makes an exhaust a valid source for AKR, if only the conditions for excitation can be satisfied. 

Since the auroral ionosphere is the canonical paradigm of an ECMI source, such conditions are in favour of the assumption that whenever reconnection in the auroral ionosphere can proceed, the exhaust region will serve as source of AKR as the result and signature of the ECMI.

The release from the noted deficiency in magnetic field strength -- and thus in cyclotron frequency -- is to assume that reconnection takes place in a so-called \emph{magnetic guide} field $\vec{B}_g$, which points \emph{along} the current flow $\vec{J}$ and is of sufficient strength $B_g\gg B_0$ to by far exceed the proper magnetic field $\nabla\times\vec{B}_0=\mu_0\vec{J}$ of the guide-field-aligned current. 

This, however, is exactly the configuration encountered in regions where the ECMI is believed to be at work, for example in Earth's auroral upper ionosphere, where the ambient geomagnetic field is a factor of $>$100 stronger than the magnetic field $B_0$ of the field-aligned current. Similarly, conditions in the strong magnetic fields in Jupiter's or Saturn's lower magnetospheres are as well promising.

The example given in Fig. \ref{fig-ecmi-two} is typical for weak or moderate guide-field simulations of reconnection. ECMI source  regions of interest here are, however, characterized by strong magnetic guide fields. Reconnection simulations in such fields are not available yet. They have not attracted any interest in the reconnection community, because reconnection is considered a mechanism for the release of magnetic energy and particle acceleration rather than radiation. Radiation, as mentioned above, is just a minor fraction of the released energy. 
The density of the reconnection-released energy is, at maximum, the total energy of the reconnecting magnetic fields $\lesssim 2|B_0|^2/\mu_0$. Because strong magnetic guide fields $\vec{B}_g$ along the current $\vec{J}$ are not affected by reconnection, this maximum is independent of the guide field strength. Concerning reconnection, the guide field effect is twofold: strong guide fields maintain the reconnection 2-D. In addition they magnetize the electrons (as also the ions) in the plane perpendicular to the guide field.

\subsection{Remarks}
At this place there is an important point to make. Guide-field simulations of reconnection in 2-D were occasionally performed with guide field strengths between $0.1\leq B_g/B_0< 2$. Though 2-D simulations with very weak guide fields are not completely unreasonable \citep[however, see the 3-D simulations reviewed by][]{karimabadi2013}, two-dimensionality breaks down when $B_g/B_0\gtrsim\frac{1}{2}$, because then the total magnetic field becomes a spiral of opening (winding) angle $\alpha\sim\tan^{-1} (B_g/B_0)$. Results of 2-D simulations become unreliable in this case, as the field is three-dimensional. On the other hand, at dominating guide fields $B_g\gtrsim 5\,B_0$, say, when the total magnetic field stretches, the stiff guide field restores two-dimensionality, which becomes robust then. This is clearly the case for reconnection in the auroral upper ionosphere with its strong geomagnetic field and the comparably much weaker fields of any field-aligned currents. 

Similar arguments apply of course also to Jupiter and other strongly magnetised planets, the sun and stars. Just for this reason reconnection has never been seriously considered to take place and become important here in the ECMI. Instead one rather invoked generation of anomalous resistivities by plasma instabilities, and even envisaged something called `breaking of field lines' which naturally would violate the basic laws of electrodynamics.   

\section{The proposed model}
Having said that much to set the framework, we are prepared to turn to the physical model settings. This model consists of the following three parts:
\begin{itemize}
\item[($a$)] the reconnection in strong guide fields, which is the basic condition for \vspace{-1ex}
\item[($b$)] the mechanism of ECMI, which generates the radiation field, and \vspace{-1ex}
\item[($c$)] the conditions for propagation and escape from the plasma.
\end{itemize}

\subsection{Requirements for reconnection}
Reconnection is based on the assumption of annihilation of oppositely directed magnetic fields, which belong to a sheet current of comparably narrow width. Such narrow current sheets exist in the auroral magnetosphere with current flowing along the strong geomagnetic field downward into the ionosphere where they close with perpendicular currents and back upward along the field in some slightly broader sheets. These currents are carried by electrons which flow upward respectively downward in the adjacent field-aligned regions. 

Magnetic field-aligned currents of this kind are pulses of kinetic respectively inertial Alfv\'en waves (KAW, IAW), depending on whether the plasma $\beta>m/m_i$ or $\beta<m/m_i$, i.e. larger or smaller than the electron-to-ion mass ratio. At high altitudes in the comparably weak geomagnetic field and high plasma temperatures these  are KAW have transverse dimensions {of the ion gyroradius $r_{gi}$}, corresponding to broader current sheets. When entering the auroral source region of AKR they become IAW of transverse scale of the order of the electron-inertial  length. This behaviour can be read from the KAW dispersion relation \citep[cf., e.g.][Eq. (10.181), revised edition 2012]{baumjohann1996}, which in the upper ionospheric IAW region becomes
\begin{equation}
\omega_\mathrm{kA}^2\approx \frac{k_\|^2V_A^2}{1+k_\perp^2\lambda_e^2}, \qquad k_\|^2\ll k_\perp^2 \sim \lambda_e^{-2}
\end{equation}
It propagates along the guide field at a fraction of the Alfv\'en speed while it belongs to a narrow current sheet of typical scale the order of $\lambda_e$. Thus the upper ionospheric field-aligned current system is structured, consisting of a series of electron inertial-scale current sheets varying with time. Each sheet carries its proper magnetic field $\vec{B}_0$. This field is perpendicular to the main geomagnetic guide field $\vec{B}_g\perp\vec{B}_0$ and has opposite direction to both sides of the current sheet. $\vec{B}_0\ll\vec{B}_g$ is a weak field, substantially weaker than the (geo)magnetic guide field along that the current flows. Such a current-total magnetic field ($\vec{J},\vec{B}_g+\vec{B}_0$)-configuration is schematically shown in Fig. \ref{fig-ecmi-two}. 

\begin{figure}[t!]
\includegraphics[width=0.5\textwidth,clip=]{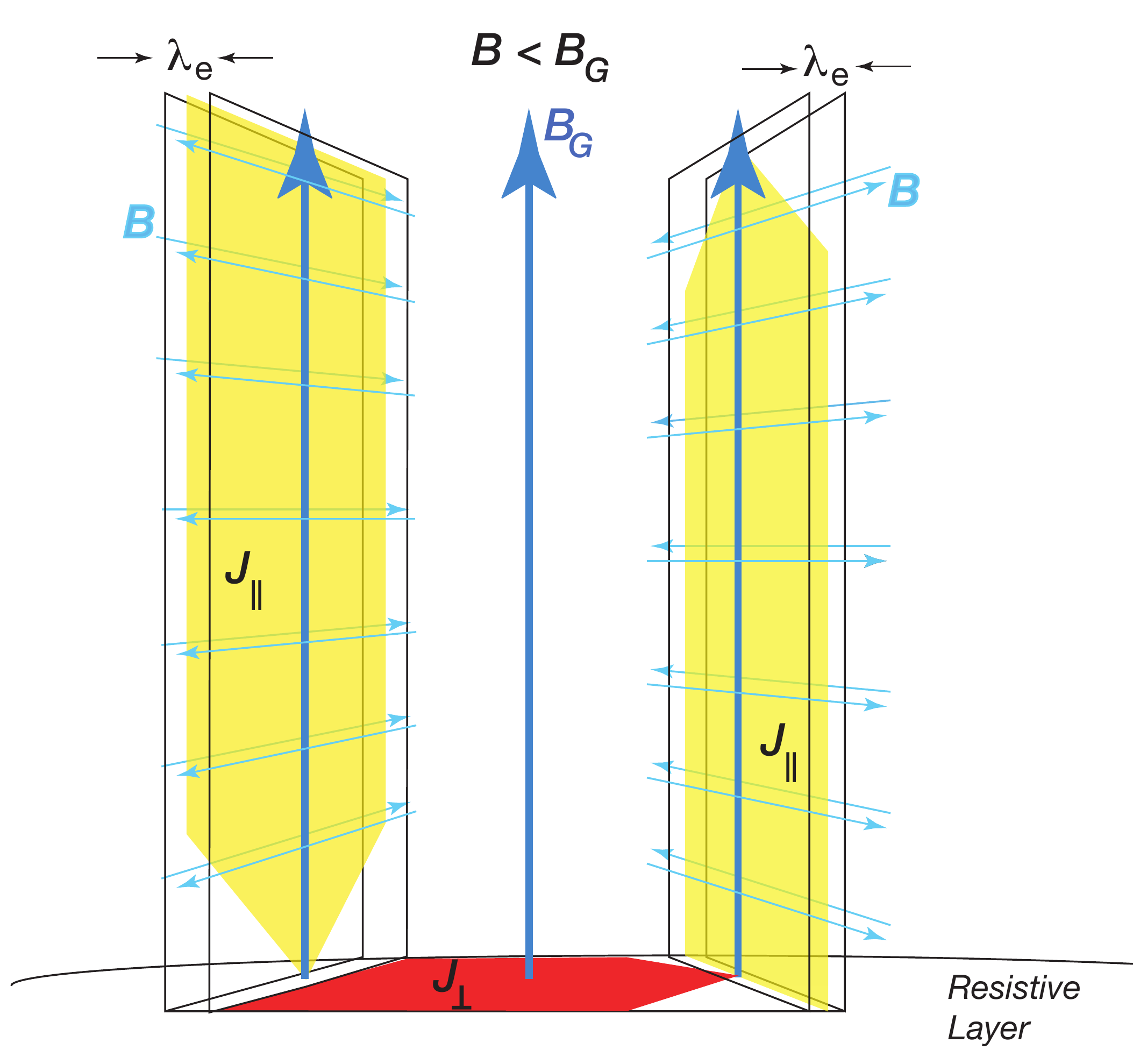} 
\caption{Two oppositely directed electron-scale current sheets with current $J_\|$ (yellow) flowing along the guide magnetic field in the case that $B_g>B_0$ is stronger than the magnetic field {$B\equiv B_0$} (light-blue vectors) of the guide-field-aligned current. The currents close via $J_\perp$ somewhere in a dissipative layer (red) where they are allowed to flow perpendicular to the field. The width of the guide-field-aligned curent sheets is assumed to be narrow, the order of the electron inertial length which would allow the current-magnetic field $B_0$ to reconnect aross the current sheet.}
\label{fig-ecmi-three}
\end{figure}

Once the current sheet becomes narrow, of the width of either the electron gyroradius or electron inertial scale, the electron magnetization decreases due to partial electron-demagnetization until it cannot anymore balance the attractive Lorentz force between the two antiparallel magnetic fields on both sides of the current sheet. Once this happens, the two oppositely directed fields collapse, and spontaneous reconnection \citep{treumann2015} sets on.  

There is a slight complication in this process, because the electrons are still magnetized in the guide field. However, this magnetization is in the plane perpendicular to the guide field. It causes a weak orbital diamagnetism along the guide field which slightly reduces the guide field but has no effect on the attractive Lorentz force. Hence, under these conditions the magnetic field of the field-aligned current carried by the IAW will collapse locally and reconnect. Locally its two oppositely directed components are perpendicular to the guide field and in the plane perpendicular to the current aross the guide field in the centre of the current sheet until they mutually annihilate, which causes an $\mathsf{X}$-point in the magnetic field of the current and the IAW to which the current belongs. The IAW develops a vertex in its magnetic field $\vec{B}_0$, the magnetic $\mathsf{X}$-point, at this location while the field-aligned current continues flowing. The IAW also continues propagating along the field, which causes a shift of the whole reconnection region, the $\mathsf{X}$-point, along the field either to higher or to lower field strengths, depending on the direction of the IAW-wave propagation. The guide field remains unaffected by this whole process of reconnection.  
\begin{figure*}[t!]
\centerline{\includegraphics[width=0.9\textwidth,clip=]{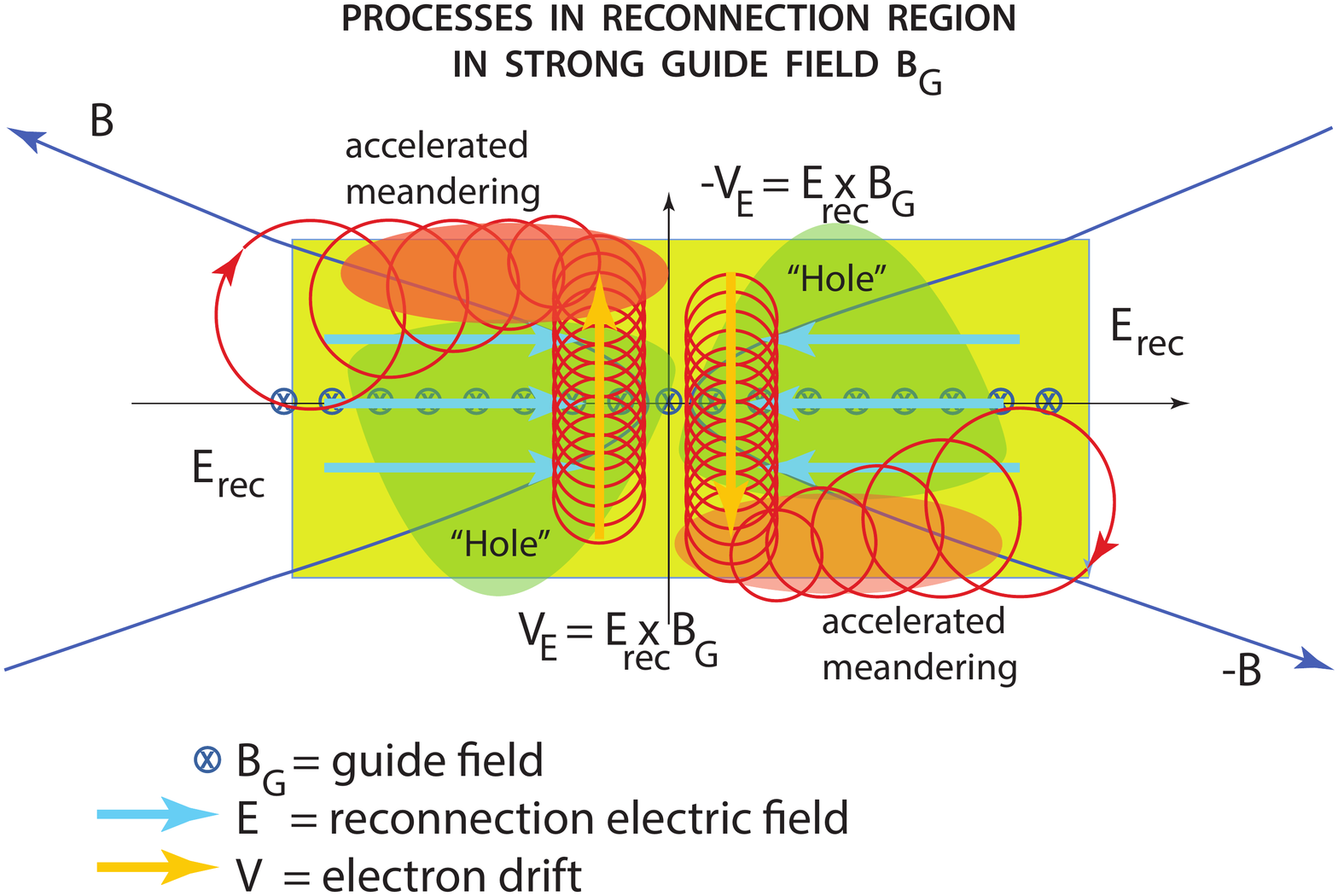}} 
\caption{Schematic of the electron dynamics inside the electron exhaust (rectangular yellow region) in presence of a strong guide field (black crossed circles) directed into the plane and antiparallel to the current. The exhaust is shown as rectangular region with reconnected magnetic field lines $B$. The blue arrows represent the reconnection electric field which points into the centre of the {\sf X}-point. Without guide field this field would expel and accelerate electrons out of the reconnection site as known from the non-guide field simulations. With strong guide field, however, it first forces the electrons to perform a $E\times B_g$ drift upward on the left, downward on the right until the electrons accumulate at the exhaust boundary where the field ceases to exist and they are reflected from the current magnetic field $B$. This forms the green electron quasi-``holes'' and the dense red regions of accumulated electrons. Reflection at the exhaust boundary forces the electrons to perform meandering motions along the guide field/current direction. During each such half-gyration they on their turning points experience the reconnection electric field and like in a synchrotron become accelerated in the direction opposite to the electric field. This causes a stepwise increasing gyroradius shown here in the projection onto the plane perpendicular to the guide field. In effect these electrons ultimately reach high energy in the direction of the exhaust perpendicular to $B_g$ causing deformation of the momentum space distribution. This is the condition required for the ECMI to become excited.} \label{fig-emci-four}
\end{figure*}

\subsection{Role of IAW}
In order to avoid confusion it becomes necessary to clarify the role of the KAW. This wave is required as the wave-carrier of the current. It possesses a guide-field-aligned electric field $E_{\|}^\mathrm{kA}=-ik_\|\phi_\|^\mathrm{kA}$ which accelerates electrons along the guide field to become the field-aligned current 
\begin{equation}
J_{e\|}^\mathrm{kA} = -\frac{ m}{2\mu_0T_e}\frac{\omega_\mathrm{kA}}{k_\|\lambda_e^2}\phi_{\|}^\mathrm{kA}=\frac{im}{2\mu_0T_e}\frac{\omega_\mathrm{kA}}{k_\|^2\lambda_e^2}E_\|^\mathrm{kA}
\end{equation}
and it is the KAW magnetic field, the magnetic field $\vec{B}_0\perp\vec{B}_g$ of the KAW current, which is \emph{perpendicular} to the strong guide field $\vec{B}_g$. In the downward current region the potential $\phi_{\|}^\mathrm{kA}>0$ is positive with upward pointing gradient, and both current and electric field point downward, accelerating electrons upward. In the adjacent upward current region the directions are opposite. Accordingly, the current-carrying kinetic Alfv\'en waves propagate downward respectively upward in these regions. 

The KAW-magnetic field $\vec{B}_0$ reconnects at some location where the transverse pressure of the magnetized electrons in the wave vanishes on the electron inertial scale $<\lambda_e$ and thus cannot anymore balance the attracting Lorentz forces between the antiparallel fields $\pm\vec{B}_0$ to both sides of the field aligned current layer (the general reason for spontaneous onset of collisionless reconnection). In Earth's upper auroral ionosphere this happens within few Earth-radii in the region where the KAW changes {from its kinetic to its inertial limit} with the inertial Alfv\'en wave (IAW) dominated by electron inertial effects. This happens at a geocentric radial location where in the strong geomagnetic guide field {the plasma-to-magnetic energy density ratio $\beta=2\mu_0NT_e/B_g^2$} changes from $\beta>m/m_i$ to $\beta<m/m_i$ \citep[cf.][Section 10.6.5]{baumjohann1996}. 

{Thus, for reconnection to occur in a strong guide-field set-up like in the case of the auroral upper ionosphere, the parallel electric field of the IAW and the related electron acceleration along the (ambient geomagnetic) guide field is important as the provider of the current source. Without its presence nothing would happen, i.e. without the IAW and its parallel electric field, whose sources must be seen at a different place, there would be no field-aligned current and thus no reconnection. Otherwise the IAW's guide-field-parallel electric field has no effect on the reconnection itself. Ultimately, though, after reconnection starts and evolves to a violent state, it is the magnetic energy of the IAW which in the reconnection process at the reconnection site is transformed into heat, entropy, particle acceleration and radiation, not the energy of the main guide field. Thus, for the IAW, reconnection is a \emph{local dissipative} process.\footnote{{This is a sensitive point. Dissipation of the wave-magnetic field implies that the IAW at the reconnection $\textsf{X}$-point breaks off. Thus, when reconnecting, the IAW decays into a wave chain. Since the parallel momentum is unaffected, both the wave chain and exhausts should continue moving along the guide field. To simulate this would be most interesting, because of the expected effects in the aurora.}}  It affects or even destroys the IAW locally over the volume of the exhaust while leaving the guide field unaffected -- at least as long as one deals with weakly relativistic plasmas. In strongly relativistic plasma the situation should become more violent \citep{bykov2012,treumann2015} because shock waves and turbulence may form.} 

{So far we mentioned only the parallel IAW electric field and its principal effect. IAWs possess substantially stronger than parallel perpendicular electric fields $E_\perp^\mathrm{kA}$. These are transverse, thus being perpendicular to both the wave and ambient (guide) magnetic fields. Since, as explained above, electrons in a strong guide field are magnetized in the plane perpendicular to the guide field, the effect of the perpendicular KAW electric field is that it causes an $E_\perp^\mathrm{kA}\times B_g$-drift of the exhaust electrons. Qualitatively, this drift is along the exhaust and thus along the reconnecting field $\vec{B}_0$. It acts amplifying and modulating the acceleration of the electrons along the exhaust. In this way it contributes to the deformation of the electron momentum distribution in the direction perpendicular to the ambient magnetic guide field. Its presence is therefore quite important for the ECMI though of minor importance for the reconnection. Also, it just adds a constant speed along the exhaust. To quantitatively infer its contribution requires numerical PIC simulations. In contrast, the weaker reconnection electric field which is along the exhaust and parallel to $\vec{B}_0$ causes a continuous acceleration of the gyrating electrons when experiencing a large number of cyclotron gyrations in the exhaust, as has been noted earlier. This is an effect that is particulary suited for investigation in simulations.}

Reconnection goes on the expense of the IAW magnetic field, i.e. the wave-carried field-aligned current. Both suffer energy losses caused by reconnection, {but these are grossly independent of the guide field, in particular in strong guide fields. It is only in weaker guide fields $B_g\sim B_0$ where the total field becomes wound up and the whole reconnection process is 3-D that, therefore, the guide field energy contributes to reconnection. In the auroral ionosphere we are confronted with the example of a strong guide field case.}\footnote{{As for a historical remark, reconnection has so far never been considered to be of any relevance for processes going on in the auroral region with its strong ambient field. It has been known for long that KAWs and IAWs play an important  role as current carriers, accelerating electrons \citep[e.g.,][]{chaston2003} via their parallel wave electric fields. KAW soliton formation has been considered as a nonlinear effect \citep[cf., e.g.,][and others]{hasegawa1976,dubinin2005}, and current-driven ion and electron holes \citep[cf.,][]{carlson1998,pottelette2005} have been identified (mainly from FAST and Viking spacecraft observations). However, the main dynamical effects were attributed to field-aligned current-generated high anomalous resistivities in the aurora \citep[cf., e.g.,][]{lotko1998} and some kind of `field-line breaking'. Anomalous resistances turned out rather inefficient, while field-line breaking is physically questionable. Reconnection in strong guide fields, if experimentally confirmed, cures most of these deficiencies, though it still requires simulational and experimental verification. Locally it generates a region of high wave-magnetic dissipation, particle acceleration, and radiation signatures in ECMI. Statistically, the dissipation could possibly even be described as a localized anomalous wave-generated resistance. However, the basic mechanism is not to be sought in any wave-nonlinearity, rather it is the by now well-established and violent reconnection process.}}

\subsection{Effect of reconnection}
In strong guide fields the effects shown in Fig. \ref{fig-ecmi-two} should with high probability evolve in a much more pronounced way, causing deep electron exhausts, large asymmetries and even stronger acceleration/heating and anisotropies of the exhaust electron component. 

Fig. \ref{fig-emci-four} gives a schematic representation of the various processes in the reconnection region with the strong guid field and electric current pointing into the plane. According to the simulations with weak guide fields the electric field caused by the spontaneous reconnection of the oppositely directed magnetic field components points inward towards the $\mathsf{X}$ point and is nearly parallel to the reconnected magnetic field. However, in the strong guide field it causes an $E_\mathit{rec}\times B_g$-drift of the magnetized electrons perpendicular to the guide and reconnection electric fields which gives rise to the two asymmetric electron depletion regions (exhausts or holes) and the electron accumulations at the boundaries of the reconnection region. The accumulated electrons cannot leave to the environment, however, being trapped at the boundary. They perform a partial meandering motion in the crossed electric and magnetic fields and become accelerated along the boundary of the reconnection site as indicated in Fig. \ref{fig-emci-four}. This kind of heating and acceleration produces momentum space anisotropies and deformations in the electron distribution function which lead to the required perpendicular momentum space gradients. 
\begin{figure}[t!]
\includegraphics[width=0.45\textwidth,clip=]{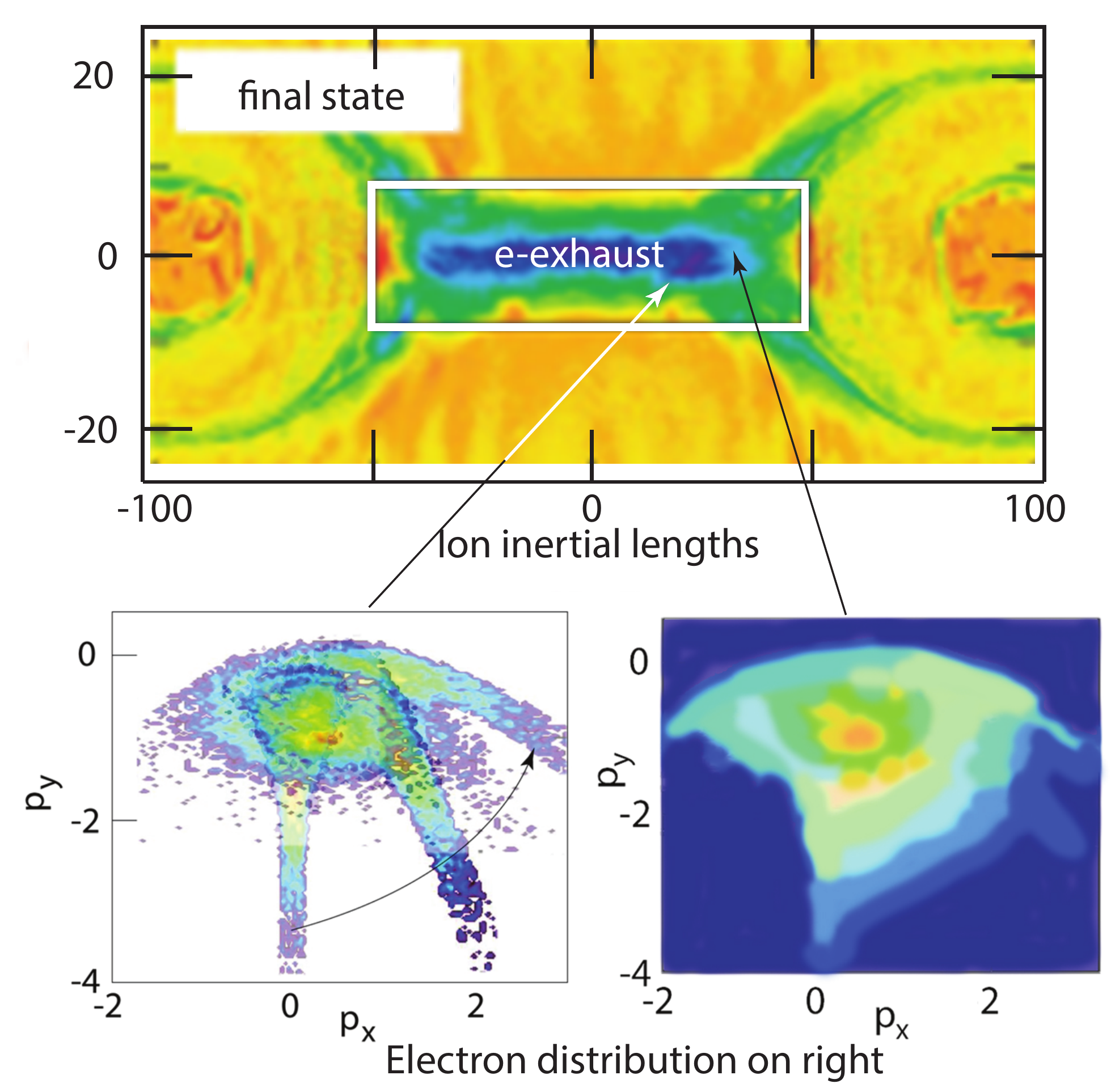} 
\caption{Simulation of particle density (\emph{top panel}) in 2d non-guide field reconnection \citep[after][]{bessho2011}. The about rectangular shape of the electron exhaust is recognized as the green domain with nearly no electrons in its centre. In this non-guide field simulation the length of the exhaust extends to roughly $50\,\lambda_i$ on both sides of {\sf X}. More interesting is the evolution of the electron distribution function (\emph{bottom panels}). The left panel is a superposition of snapshots of distribution functions at increasing distance to the left from the {\sf X} point. The distribution gradually evolves into the average distribution shown on the right and exhibits a strong anisotropy in momentum space. The light green ring with its yellow maximum shows the perpendicular anisotropy forming a well-expressed ring structure (the quantitative color scale is suppressed). In the presence of a guide field the distribution forms a ring perpendicular to the guide field. Without guide field such distributions excite the Weibel instability \citep{weibel1959,fried1959,yoon1987,treumann2012a,treumann2014} and create a secondary magnetic field which structures the current and leads to magnetic turbulence. In presence of a guide field Weibel instability is suppressed while the anisotropy favours the ECMI.} \label{fig-emci-three}
\end{figure}

Deformation of the initially isotropic electron distribution in the process of reconnection has recently been shown in PIC simulations without the presence of any guide field. 

An example of this is reproduced in Fig. \ref{fig-emci-three}, adapted from \citet{bessho2011}. It shows the rectangular exhaust region developing in absence of guide fields as a symmetric region extended along the reconnecting magnetic field $\vec{B}_0$. The enormous dilution of plasma in this region is striking. Nearly all the electrons are expelled in the course of the reconnection from the $\mathsf{X}$-point, an effect we already saw in its asymmetric form in the weak guide field simulation. 

The lower two panels show on the left three mutually overlaid snapshots of the momentum space electron distribution function at three different locations in the exhaust during the evolution of the reconnection. It is clear that the distribution function develops steep gradients mainly in $p_y$. The ultimate distribution at the right end of the exhaust is seen in the right panel. It exhibits an energetic ring distribution which strongly resembles the model distributions used in the electron hole ECMI case \citep{treumann2011,treumann2012}. 

Such a momentum distribution possesses moderate if not steep gradients in the two perpendicular momenta \emph{perpendicular to the current} (and thus to the guide field). Moreover, the ECMI resonance curve in momentum space is in this like the hole case a large section of a circle along the yellowish maximum in the distribution. It satisfies the necessary condition for excitation of the ECMI under conditions of reconnection.  The main difference between these simulations and those with guide field will be found in a much stronger asymmetry in the distribution function and therefore a deformation of the resonance curve. However, we may in this model assume that for the principal effect these asymmetries just represent technical complications in performing calculations.

\begin{figure}[t!]
\centerline{\includegraphics[width=0.5\textwidth,clip=]{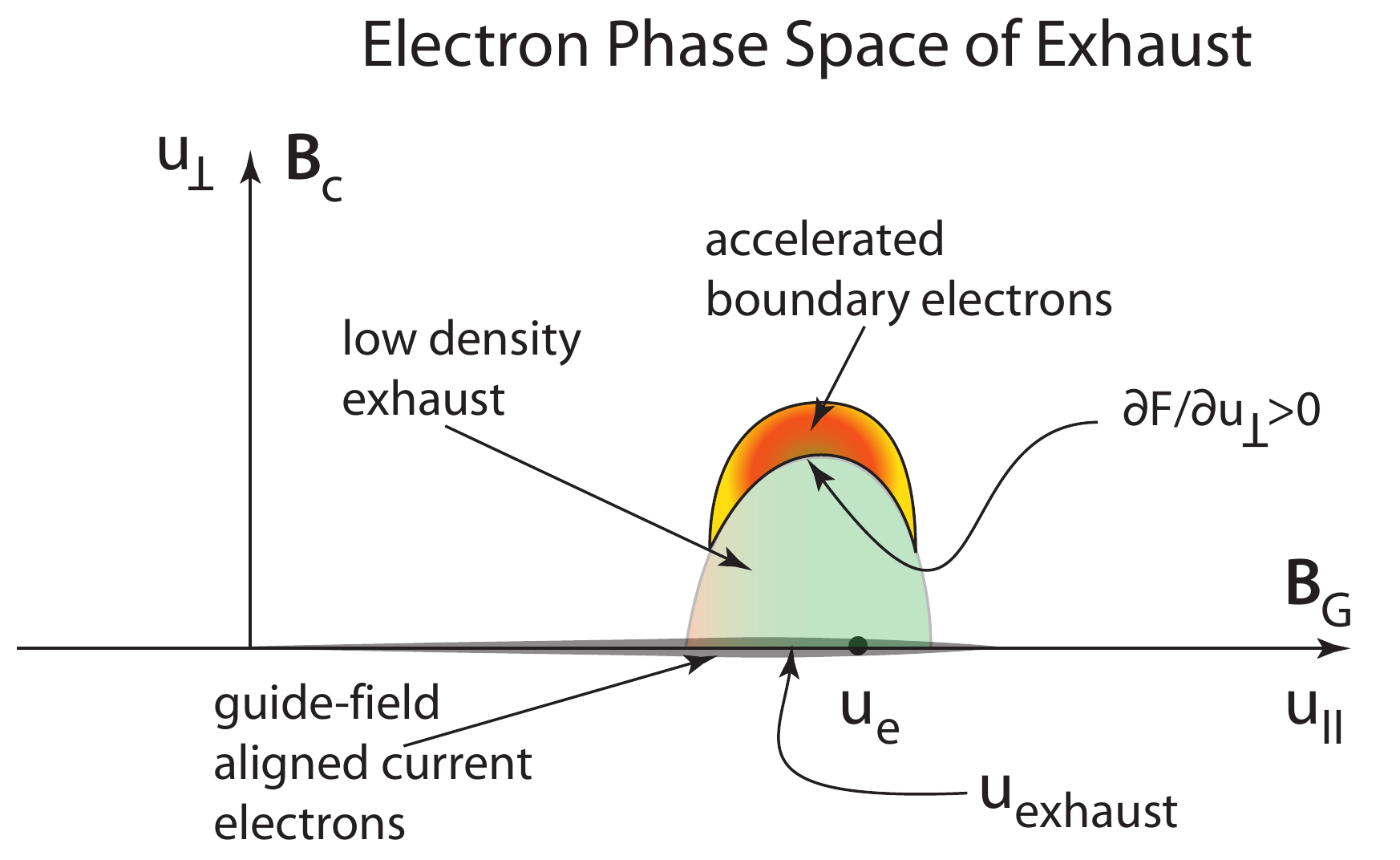}} 
\caption{{Mass-normalised momentum-phase-space of electrons in strong guide field exhaust with $\vec{u}=\vec{p}/m$.} The guide-field-aligned current electrons have a broad electron distribution (similar to those observed in the AKR downward current region) shown here along the $u_\|$ axis which is also the direction of the guide field $\vec{B}_g$. The greenish region is the highly electron diluted physe space domain of exhaust electrons. This region is filled with perpendicular drifting dilute electrons which move across the hole and magnetic fields by $E\times B_g$ drift to the boundaries. The banana shaped colored region is filled with the exhaust-electric field boundary electrons which form non-symmetric beams at the boundary of the exhaust which are directed parallel/antiparallel (depending on the part above or below the exhaust {\sf X}-point) to the reconnected field-aligned current magnetic field $\vec{B}_0$. It gives rise to a perpendicular phase-space distribution gradient $\partial F(\vec{u})/\partial u_\perp >0$ which is capable of exciting the ECMI.} \label{fig-emci-five}
\end{figure}

\subsection{Exhaust properties of relevance for the ECMI 2}
With these considerations and clarifications we have arrived at the point, where we can summarize the effects of reconnection which are of relevance to the ECMI as follows:
\begin{itemize}
\item[($a$)] Reconnection, whether or not proceeding in guide fields, generates extended highly electron-diluted `exhaust' regions, which are elongated in the direction of the reconnecting magnetic fields. With guide fields present, these exhausts split into two asymmetrically located regions of diluted electron populations. In addition, complementary regions of electron accumulations evolve at the boundaries of the reconnection site. \vspace{-1ex}
\item[($b$)] The electron momentum distribution functions in and at the boundaries of the exhaust become accelerated in a combination of the electric field that is generated in the process of reconnection, and the cross-field drift of the electrons in the strong guide field. This combined action causes a deformation of the electron momentum space distribution in the plane perpendicular to the guide field. It generates sufficiently steep perpendicular electron-momentum space gradients $\partial f_e(\vec{p}_\perp)/\partial{\vec{p}_\perp}$ on the electron distribution function. These gradients exist in both perpendicular components of the electron momentum, being the requirement for the ECMI. Acceleration of the electrons in reconnection in general, whether in guide fields or not, warrants the at least weakly relativistic nature of the electron distribution. \vspace{-1ex}
\item[($c$)] The extension of the reconnection site and the electron exhaust along the reconnection field and thus perpendicular to the guide field amount to several ion-inertial lengths which corresponds to roughly 100 $\lambda_e$ or up to $10^3\ \lambda_D$. Compared to electron holes, reconnection exhausts are therefore macroscopic objects. Their extension along the current and guide field cannot be read from any 2-D simulations but is at least of the order of few ion-inertial lengths. It will not exceed the parallel wavelength $\lambda_\|\gg\lambda_e$ of an IAW pulse propagating along the guide field. Hence exhausts are 3-D objects of extensions of few to several $\lambda_e$ in each spatial direction and can have further substructure.\vspace{-1ex}
\item[($d$)] As for a final observation one may note that the relation between the generation of electron exhausts and IAWs implies that an electron exhaust, once formed in reconnection, will necessarily propagate along the guide field together with the IAW with speed of a substantial fraction of the Alfv\'en velocity $V_A$ based on the ambient plasma parameters. Since the IAW carries the current, it propagates downward in the downward current region and upward in the upward current region. One therefore expects  similar propagation directions for the exhausts.
\end{itemize}

\section{Application to the ECMI}
After these preparatory clarifications of our model we are now in the position to apply the reconnection paradigm to the ECMI. In fact, the similarity is striking between an electron exhaust in reconnection and an electron hole produced by the nonlinear evolution of the Buneman instability in unmagnetized as well as magnetized plasmas. Both are localized electron depletions; both contain electron distributions with perpendicular momentum space gradients and, under certain conditions, weakly relativistic electrons; both propagate along with the current flow. A schematic of the structure of the exhaust in the normalized relativistic momentum space $\vec{u}=\vec{p}/m$ is given in Fig. \ref{fig-emci-five}. 

The main differences between the model of an electron hole and the exhaust are the different sizes. Electron holes are Debye-scale entities, while reconnection exhausts are comparably very large electron inertial-scale entities. This makes the latter indeed interesting candidates for hosting the ECMI. Moreover, this similarity allows for a direct transfer of our ECMI theory of radiation produced by electron holes to the ECMI radiation which can be expected of being generated in a reconnection electron exhaust -- with the exhaust playing the role of an ``electron quasi-hole''.

\subsection{ECMI from electron holes/exhausts -- a briefing}
Let us recall the properties of the ECMI in the case of electron holes \citep{treumann2011} and write it in terms of the reconnection parameters. For simplicity assume a circular hole.\footnote{The geometrical form of the exhaust and thus the resonance curve could be modelled as ellipses which would introduce excentricity as another free parameter thus complicating the calculation a bit. Though this is not too difficult to manage mathematically, we ignore this complication because it does not provide any deeper insight into the physics of the role reconnection plays in the ECMI.} Working in the hole frame, the case also natural in exhausts, the resonance condition in phase space is a circular section in perpendicular momentum space of radius
\begin{equation}
R_{res}=\sqrt{2(1-\nu_{c})}, \quad \nu_{c}=\omega/\omega_{c}
\end{equation}
where now, because of the strong guide field $\vec{B}_g$, 
\begin{equation}
\omega_{c}=\frac{eB_g}{m}\sqrt{1+\frac{\vec{B}_0^2}{\vec{B}_g^2}}\approx \omega_{cg}\left(1+\frac{1}{2}\frac{B_0^2}{B^2_g}\right)
\end{equation}
The maximum growth rate is obtained at maximum resonance
\begin{equation}\label{rm}
R_m\approx U-(\Delta u)^2/U, \quad U=V_A/c,~\Delta u=v_e/c
\end{equation}
and $V_A$ the drift velocity of the hole/exhaust, which in our case is the Alfv\'en speed or a larger fraction of it based on the guide field,  and $v_e$ is the thermal speed of the hot trapped electrons, which is small with respect to $c$ but by no means small compared with the Alfv\'en speed. Then $\Delta u\sim 0.1$, and the second term in $R_m$ can be dropped. Moreover, we have approximately
\begin{equation}
R_m\approx U=\sqrt{\frac{m}{m_i}}\frac{\omega_{cg}}{\omega_e}
\end{equation}
The maximum normalized growth rate is obtained when using the full expression (\ref{rm}) for $R_m$ in the growth rate \citep[cf.,][]{treumann2011}, which then becomes
\begin{equation}\label{eq-growthrate}
\frac{\Im\,(\omega)}{\omega_{cg}}\bigg|_{R_m}\approx\ \frac{\alpha\pi^3}{4\nu_c}\frac{\omega_{e}^2}{\omega_{c}^2}\bigg(\frac{V_A}{v_e}\bigg)\bigg(\frac{c}{v_e}\bigg)\exp\bigg(-\frac{1}{2}\frac{v_e^2}{V_A^2}\bigg)
\end{equation}
where $\alpha \approx  N_{h}/N_e$ is the ratio of the exhaust density at the depletion of electrons to the density of the environment, a number of order $\alpha\lesssim$ O\,(0.1) or so. This growth rate is inversely proportional to the normalized frequency $\nu_c$ and thus decreases algebraically with frequency (and the harmonics as well). In strong guide fields the contribution of $B_0$ is negligible, and the cyclotron frequency is solely determined by the guide field $B_g$. {The argument of the exponential is $v_e^2/2V_A^2=\beta (m_i/m)$. In the IAW region where, according to our discussion, reconnection becomes spontaneous and thus violent, $\beta<m/m_i$, and for the ambient plasma the exponential is between 1 and $\sim1/\mathrm{e}$. However, the hot interior electrons have higher thermal speed. They enter the growth rate, increase the interior $\beta$  and make $V_A\lesssim v_e$, with the exponential well compensating for the factor $c/v_e\sim 30$. The normalized growth rate is therefore small. It must be less than the cylotron frequency. If we assume for the exponential $<3\times10^{-4}$, we have the interior electron thermal to Alfv\'en speed ratio $v_e/V_A\gtrsim4$, which yields}
\begin{equation}
\Im(\omega)\lesssim 0.02\frac{\omega_{cg}}{\nu_c}\frac{\omega_e^2}{\omega_c^2}
\end{equation}
This is determined by the ambient ratio $\omega_e^2/\omega_c^2$  thus being a substantial fraction of $\omega_{cg}$ with fundamental emission close to and slightly below $\omega_{cg}$. In the above cited paper we estimated that the frequency width below $\omega_c$ is of the order of
$\Delta\nu_c\sim 2\beta_e^2$. Taking $\beta_e\lesssim m/m_i$ (as required for the IAW) implies that the difference between the guide field cyclotron and the maximum emission frequencies at the fundamental X-mode is of the order of $\Delta\nu_{cg}\sim 10^{-3}$, a small number. At $\sim 300$ kHz in the AKR this difference corresponds to $\sim 300$ Hz. Emission is practically at the guide-field cyclotron frequency.  Since the cyclotron period is short, there is plenty of time to grow for the wave. The fundamental will be trapped inside the exhaust for long time, while higher harmonics can possibly escape, depending on the ambient cyclotron to plasma frequency ratio.

{Trapping of radiation inside the exhaust implies a number of other interesting effects which have been discussed in relation to electron holes \citep{treumann2011, treumann2012}. These include the possibility for amplification due to sufficient time for experiencing many growth cycles, quasilinear saturation of the radiation intensity, depletion of the perpendicular momentum gradient in the electron distribution, interferences, as well as other nonlinear effects at large radiation amplitude. One may expect relatively high saturation levels of the trapped modes. Trapping also implies convective radiation transport in the exhaust along the guide field to locations where the plasma conditions become favourable for radiation release.}

\subsection{Spatial displacement of ECMI source}
{The above growth rate also exhibits another observationally important effect. It is restricted to the location of the exhaust in space. However, the exhaust moves in two directions. Firstly it moves with the field-aligned current along the guide-magnetic field, an effect which is included in the Alfv\'en speed $V_A$ and in the cyclotron frequency $\omega_c$. It also moves at a slowlier speed perpendicular to the guide field shifting away from the reconnection $\mathsf{X}$-point to either left or right along the reconnection magnetic field (i.e. along the field-aligned current-sheet plane), an effect which, when compared with $V_A$  probably causes a weak modulation only. The main modulation of the growth rate is caused by the spatial dependence of the Alfv\'en speed along the guide field, when the $\mathsf{X}$-point moves with the Alfv\'en wave along the field. The frequency dependence on location and thus magnetic field is contained in $\omega_c\sim  B_g$, while $\nu_c\sim 1$. Thus the variation of the growth rate with coordinate $s$ along the guide field is determined essentially by the spatial variation of $B_g(s) /\sqrt{N(s)}$ along the guide field as 
\begin{equation}
\Im(\nu_c, s)\sim\sqrt{N(s)}/B_g(s)
\end{equation}
In case $B(s)/N(s)\approx$ const, the growth rate behaves like $\Im(\nu_c,s)\sim 1/\sqrt{N(s)}$. In addition the exponential factor affets the growth rate if the Alfv\'en speed reduces, which happens only when the guide field becomes very weak. Emission very close to the cyclotron frequency or its harmonics will thus \emph{intensify with decreasing density}. This should be the situation in the upper auroral ionosphere for outward propagation.}

\section{Discussion}
The first condition for the ECMI is that there is more magnetic energy per electron in the volume of the exhaust than the rest mass energy of an electron:
\begin{equation}
\frac{\omega_e^2}{\omega_{ce}^2} =\frac{m_ec^2}{B^2/\mu_0N} <1
\end{equation}
In this case it is clear that, as mentioned before, there are not enough electrons to absorb any excess in free energy and carry it away in order to establish thermodynamic equilibrium. In order to reduce the accumulated free energy in the perpendicular velocity space gradient  the plasma reacts by directly exciting the free-space waves. In our case of reconnection the free energy is the energy of the current, i.e. of the IAW. The process which dissipates the energy is spontaneous collisionless reconnection. Radiation eliminates just a fraction of the available reconnection energy. The main dissipation proceeds via plasma heating and acceleration in reconnection. Nevertheless, the conditions are in favour of direct excitation of radiation via the ECMI. Thus, reconnection in strong guide fields will always be accompanied by radio emission which is caused by the ECMI in the extended electron exhaust. Its observation from remote should provide information on the plasma state (strength of magnetic field, density, plasma processes, presence of reconnection, field-aligned currents, electron beams etc.).

\begin{figure}[t!]
\centerline{\includegraphics[width=0.5\textwidth,clip=]{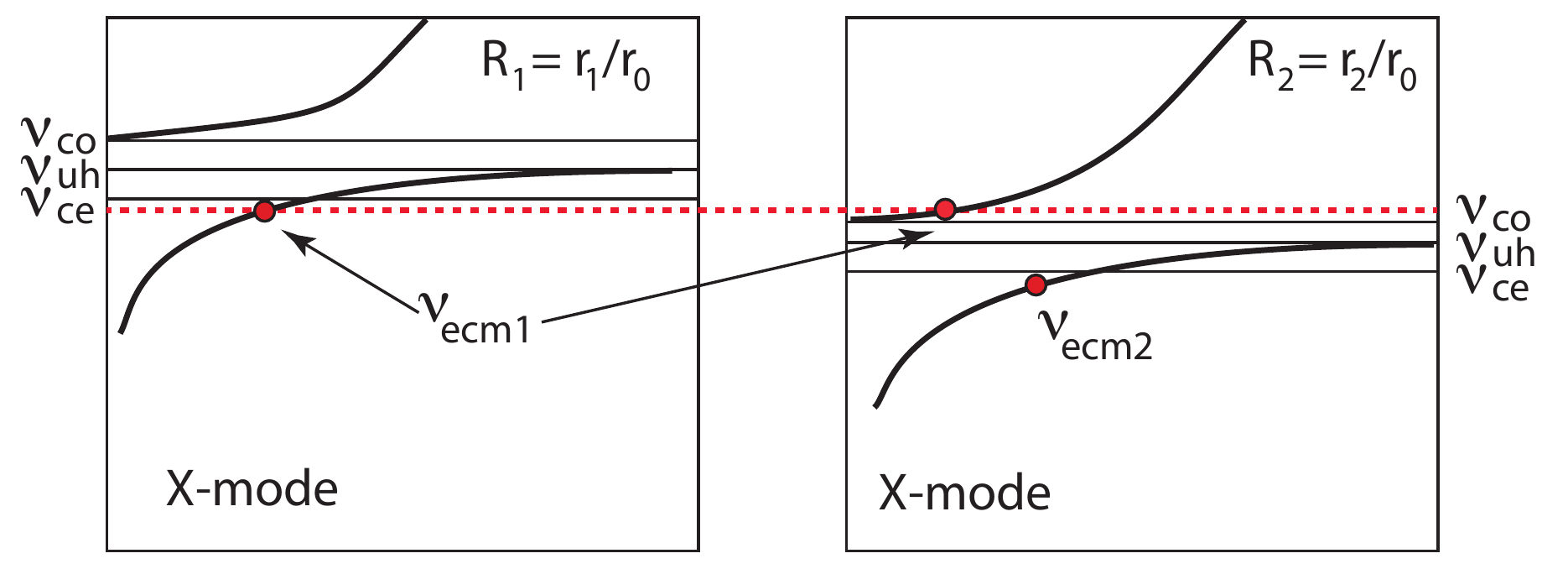}} 
\caption{Illustration of the requirements for escape of ECMI radiation excited in the X mode to free space if the radiation is generated a frequency beneath the electron cyclotron frequency of the guide field and moves from $R_1 \to R_2$ with $r_0$ distance of normalization. Excitation is on the left at a frequency shown as the red dot. It occurs on the lower branch of the X mode. In a magnetic guide field of spatially decreasing strength (e.g. a dipole field) from location $r_1<r_2$ to $r_2$, the cut off (index co), upper hybrid (uh) and cyclotron frequencies (ce) drop below the emission frequency $\nu_{emc1}$ at $r_1$. If the wave is able to tunnel the gap on the way to the new location, its frequency $\nu_{emci1}$ will be found on the free space branch of the X mode and can escape. Tunnelling is non-trivial. It implies a rather severe condition on the radiated wavelength, angle and density gradient. At $r_2>r_1$ emission is at the low indicated frequency $\nu_{emc2}<\nu_{emc1}$ and is trapped in the plasma. An outmoving source will emit downmoving tones. No upmoving tones are seen from remote. They are inhibited by the condition of escape but can occur inside the plasma. } \label{fig-emci-six}
\end{figure}

At its fundamental the ECMI generates emission below but very close to $\omega_{c}$ based on $B_g$. Because the emission generated at location $r_1$ cannot escape, the source where the emission is trapped, must first move to a location where the radiation frequency $\nu_{c}(r_1) >\nu_{Xc}(r_2)$ exceeds the X mode cut-off frequency, as illustrated in Figure \ref{fig-emci-six}.
For an external observer any narrow frequency signal moving down or up in frequency maps the local spatial variation of $B_g$ that is related to the spatial displacement of the electron exhaust along $\vec{B}_g$ to lower or higher magnetic fields. The radiation will always be cut-off once the emission frequency drops below the local X mode cut-off (assuming it is generated in the X mode almost perpendicular to the local field). Changing emission frequency will thus have to be attributed to a motion of the exhaust, which is restricted mainly to the direction along the field and thus a consequence of the velocity of the IAW. 

An observer inside the plasma will be unable to locate the radiation source unless a measurement of the direction becomes possible. For such an observer (spacecraft) any narrowband emission signals can both rise and fall in frequency. Observed from remote, however, falling tones map the global magnetic field variation with distance. Rising tones may be attributed to Doppler-shift effects of an approaching source. Broad emission bands in frequency imply that many emission sites exist simultaneously in an extended source containing a spatially variable magnetic guide field such that many places emit almost simultaneously. Such a band will always exhibit a low frequency cut-off implied by the escape conditions.

What concerns emissions observed from the Sun, the ECMI model of reconnection offers a wide field of possible applications. The solar atmosphere and corona contain very strong magnetic guide fields, are subject to field aligned current flow which are either carried simply by electron beams or also by kinetic Alfv\'en waves. These are restricted to narrow flux tubes and can reconnect across the ambient guide magnetic rield. Such processes may lead to all kinds of emissions in the ECMI and will always indicate reconnection to go on.  

\subsection{Caveats and unresolved problems}

The remaining problems concern the confinement of the radiation and its escape to free space which is required for remote observations.\footnote{{These considerations are based on cold plasma theory of the electromagnetic modes, the simplest case. It is clear that the dispersion of all waves changes when the electron plasma becomes hot. What concerns the KAW, so temperature increase implies larger gyroradii thus favouring the transition to the inertial range. This means that the condition $\beta<m/m_i$ is relaxed, and reconnection may set on already in the KAW regime, in the auroral ionosphere at higher altitudes and weaker magnetic fields. Higher temperatures do also relax the escape conditions for the fundamental which in reality might thus not be as severe as they appear in our discussion. Higher harmonics, which are also excited by the ECMI though at decreasing growth rate, should anyway have less problems to escape to the environment and free space. }} There is no problem for the O-mode to escape if excited. The fundamental X mode, on the other hand, cannot escape for the obvious reason that it is excited by the ECMI at frequency $\omega<\omega_{ce}$. Even in the case of the guide field this condition is a severe restriction on the emission. If the guide field is strong, as is the case close to a strongly magnetized planet, the X mode may escape from the exhaust to the surrounding plasma but cannot leave from it to free space unless it has moved out to a region where its frequency exceeds the X mode cut-off $\omega_{Xc}$, i.e. roughly the local upper hybrid frequency. Still a problem remains as it has to bridge the no-propagation gap between the upper hybrid  and cut-off $\omega_{uh}\leq\omega\leq\omega_{Xc}$  frequencies. This requires sufficiently steep density gradients roughly sharper than one wavelength which is, however, a weak restriction only when the field is very strong and the density low. With $B_0$ the magnetic field at excitation of the ECMI and $B_{esc}$ the field strength at escape, the X mode escapes if the ratio of plasma to cyclotron frequency satisfies the condition
\begin{equation}
\frac{\omega_e}{\omega_{c,0}}< \left(\frac{B_{esc}}{B_0}\right)^\frac{1}{2}
\end{equation}
This bridging condition might not be too severe. It can be roughly reduced to a condition on the emitted X mode wavelength $\lambda_X$ at escape
\begin{equation}
\lambda_X> \pi\left.\left(\frac{\mathrm{d}x}{\mathrm{d}\ln N(x)}\right)\right|_{esc}
\end{equation}
Here it is assumed that the attenuation of the radiation intensity when crossing the stop-band drops much less than 1/e. A change of density $N$ by one order of magnitude then requires that it occurs over a length shorter than $\Delta x\lesssim2\lambda_X/\pi$. 

More interesting than these well-known problems, which apply quite generally to ECMI emission of radiation (but see the last footnote 5), is that the exhaust is not fixed to a spatial location. Because it is a feature that happens to take place in the IAW which propagates at the rather high Alfv\'en velocity along the strong guide magnetic field, the exhaust propagates along the magnetic guide field. Any slow deviation from the field direction caused by the non-vanishing perpendicular wave speed, a fraction of $V_A$, is interesting though of little importance for the radiation. It may, however contribute to direction induced modulation and also Doppler shift in the emitted frequency depending on the relative locations of radiation source (exhaust) and observer (spacecraft, rocket). Effects of this kind have been reported as typical for the AKR fine structure.

\subsection{Conclusions}
The ECMI has been discovered in auroral physics \citep{wu1979} and was proposed to explain the occurrence of the surprisingly intense terrestrial kilometric radiation \citep{gurnett1974}, later dubbed AKR and found to be related to the occurrence of strong magnetic field-aligned currents, the generic case of application of our theory. 

Reconnection has so far not been brought into discussion in relation to the ECMI. Instead, for long time there was talk going on referring to `breaking of field lines' in order to generate sufficient anomalous resistance along the strong auroral geomagnetic field. In fact, as our model suggests, there is no breaking of field lines. Field lines are an idealization of magnetic flux tubes on scales exeeding the average electron gyroradius which carry a magnetic flux 
\begin{equation}
\Phi= \int \vec{B}\cdot\mathrm{d}\vec{F} \approx \pi \langle r_{ce}\rangle^2B, \quad \langle r_{ce}\rangle= \frac{m_e}{e}\frac{\langle v_{e\perp}\rangle}{B} 
\end{equation}
where  $\langle v_{e\perp}\rangle$ is the ensemble average of the perpendicular velocity of electrons, and $\langle r_{ce}\rangle$ is the corresponding electron gyroradius. 
Reconnection is the process of rearranging this flux on the scale \emph{below} the electron gyroradius. This means that on those scales identification of field lines at plasma temperatures makes no sense (at temperatures far below plasma, field lines are defined as flux tubes containing one elementary quantum flux $\Phi_0=\hbar/e$. These do never ever break but are exchanged in integer numbers). 

Hence, instead of field-line breaking, the process that {presumably} takes place is reconnection in the magnetic field of the field-aligned current respectively the corresponding IAW that transports this current while the guide field remains completely unaffected. In this process the energy stored in the field-aligned current (carried by the IAW) is consumed, by no means is the guide field energy affected. The consumed energy is converted into generation of the exhaust, the reconnection electric field, which points perpendicular to the guide field, an $E_{rec}\times B_g$-drift of electrons, electron synchrotron acceleration when meandering, and ultimately excitation of the ECMI in the exhaust. {Unfortunately, no measurements of the magnetic dissipation rate in the IAW have, at least to our knowledge, been made nor are available.}\footnote{{We thank the referees for bringing up this important point. Though it would be highly desirable to directly measure the dissipation rate in IAWs as it would provide information about the possibility for reconnection in a real system like Earth's upper auroral ionosphere, it is hard to imagine how such a measurement could be performed. One way would be to think of tracing back the observation of IAW acelerated electrons \citep[as, for instance, observed by][]{chaston2003} to the action of the reconnection-generated electric field. This, however, implies reference to so far not performed PIC simulations of strong guide field reconnection with KAW/IAW.}} 

Recognition of this process as probably being fundamental to field-aligned current systems in strong magnetic (guide) fields clarifies and rounds up the picture of generation of non-thermal radio radiation by the ECMI. It identifies spontaneous magnetic reconnection as the process that takes place not only in weak fields but also in strong guide fields whenever a field-aligned current of width on the transverse electron inertial scale flows along the guide field. This process of spontaneous reconnection in strong current-aligned guide fields is particularly violent in collisionless plasmas. 

From this point of view one may realise that collisionless reconnection in guide fields, whether weak or strong, is probably the best place for application of the `electron-hole model' for the excitation of the ECMI. The weak guide field model applies to almost all cases of reconnection going on in nature, in particular also to magnetic turbulence which is believed to dissipate via collisionless reconnection. The exceptions are ideally symmetric non-guide field counterstreaming magnetised plasmas/magnetic fields, and also the very strong guide fields carrying field aligned currents as investigated here. The latter will be restricted to strongly magnetised stars, planets, and other strongly magnetised astrophysical objects, where they should lead to very high radiation intensities. If this turns out to be true, then the combination of collisionless reconnection and ECMI is a quite general excitation mechanism of nonthermal radio radiation in the universe.

\section{Summary and final remarks}
The present paper suggests such a mechanism as a working plan for an efficient and realistic model of the ECMI under conditions of strong current-aligned guide fields in collisionless reconnection or, otherwise, collisionless reconnection in field-aligned current regions along strong ambient magnetic fields. No attempt was made to develop a full quantitative theory as this requires the performance of 2-D numerical PIC simulations under conditions of very strong guide fields $B_g\gg B_0$, determination of the electron exhaust region, fraction of plasma dilution, and -- most  important -- the microscopic weakly relativistic momentum space distribution of the hot accelerated diluted trapped electron component in the exhaust. The perpendicular momentum space gradients of this distribution are the free energy source for the excitation of the ECMI at the fundamental and low higher harmonics (and probably also the transverse electromagnetic electron Bernstein modes). These results, which are obtained from the reconnection simulations, should be used in the ECMI theory to numerically calculate the growth rates for the cases of interest, based on realistic input parameters. Performing this large program lies outside the presently proposed new model of the ECMI in strong magnetic guide fields.

This mechanism combines two well established plasma processes: reconnection and ECMI, is a local process working in comparably small regions which explains the fine structure of radiation, and is of wide applicability. In strong magnetic (guide) fields like planetary magnetospheres, the sun, stars and strongly magnetized astrophysical objects it may explain the generation of drifting, highly time variable radio emission which maps the local magnetic field strength along the drift path of the electron exhausts and provides inferences about the plasma conditions and dynamics. In weak guide fields the same mechanism should work causing low frequency cyclotron radio emission which contributes to energy loss. This is of substantial interest also in large-scale magnetic/magnetohydrodynamic turbulence where the ECMI radiation at the low electron cyclotron harmonics adds up from myriads of small-scale reconnection sites to generate a stationary radio glow of the entire turbulent region thereby bringing such large-scale turbulent regions into radio-visibility, an so far not investigated effect of considerable interest in astrophysical application.  
 
In summary, ECMI in the electron exhaust/reconnection {\sf X}-point region is a promising and interesting phenomenon that may be realized quite frequently in space and under cosmical conditions. When developing in strong guide field reconnection related to field-aligned currents the emitted ECMI radiation may become quite intense results from localized drifting reconnection regions in space.

\begin{acknowledgement}
This work was part of a Visiting Scientist Programme at ISSI Bern. We thank the ISSI Directorate for interest in this endeavour. We also thank  Raymond Pottelette for cooperation in the electron-hole radiation model, the friendly ISSI staff for its hospitality,  the ISSI computer administrator S. Saliba for technical support, and the librarians Andrea Fischer and Irmela Schweizer for access to the library and literature. The very constructive comments of the referees are gratefully acknowledged as they were of substantial help in clarifying the proposed concepts. 
\end{acknowledgement}

\newpage

\end{document}